\documentclass[ aps
               ,prl
               ,floatfix
	            ,footinbib
	            ,a4paper
	            ,superscriptaddress
               ,preprint,endfloats
               ]{revtex4}

\usepackage{rotating}
\usepackage{amsmath}
\usepackage{graphics}
\usepackage{textcomp}

\begin{document}

\title{Significant elastic anisotropy in Ti$_{1-x}$Al$_x$N alloys}

\author{Ferenc Tasn\'adi}
\affiliation{Department of Physics, Chemistry and Biology
(IFM), Link\"oping University, SE-581 83 Link\"oping, Sweden}
\email{tasnadi@ifm.liu.se}
\author{Igor A. Abrikosov}
\affiliation{Department of Physics, Chemistry and Biology
(IFM), Link\"oping University, SE-581 83 Link\"oping, Sweden}
\author{Lina Rogstr\"om}
\affiliation{Department of Physics, Chemistry and Biology
(IFM), Link\"oping University, SE-581 83 Link\"oping, Sweden}
\author{Jonathan Almer}
\affiliation{Advanced Photon Source, Argonne National Laboratory}
\author{Mats P. Johansson}
\affiliation{Department of Physics, Chemistry and Biology
(IFM), Link\"oping University, SE-581 83 Link\"oping, Sweden}
\author{Magnus Od\'en}
\affiliation{Department of Physics, Chemistry and Biology
(IFM), Link\"oping University, SE-581 83 Link\"oping, Sweden}

\date{\today}
\begin{abstract}
Strong compositional-dependent elastic properties have been observed
theoretically and experimentally in Ti$_{1-x}$Al$_x$ N alloys. The elastic
constant, C$_{11}$, changes by more than 50\% depending on the Al-content.
Increasing the Al-content weakens the average bond strength in the local
octahedral arrangements resulting in a more compliant material. On the
other hand, it enhances the directional (covalent) nature of the nearest
neighbor bonds that results in greater elastic anisotropy and higher sound
velocities.  The strong dependence of the elastic properties on the Al-content
offers new insight into the detailed understanding of the spinodal decomposition
and age hardening in Ti$_{1-x}$Al$_x$N alloys.
\end{abstract}
\maketitle
%
TiAlN coatings are today used as wear protection, for example
cutting tools. The alloy has cubic B1 structure and shows a combination of good
oxidation resistance and superior mechanical properties at elevated
temperatures as compared to TiN \cite{Horling2005}. The good high temperature behavior is
correlated to an isostructural spinodal decomposition of the Ti$_{1-x}$Al$_x$N
solid solution into coherent cubic Al- and Ti-enriched Ti$_{1-x}$Al$_x$N domains
that results in a hardness enhancement, i.e. age hardening, at 800-1000\,\textcelsius\ 
\cite{Horling2002,Knutsson2008,Mayrhofer2003}.
Several subsequent studies have highlighted different aspects of importance
for the understanding of the decomposition and stability of this materials
system, such as the influence of nitrogen off-stoichiometry \cite{Alling2008}, pressure \cite{Alling2009},
and domain growth behavior \cite{Oden2009}.

Although the phase stability and performance of Ti$_{1-x}$Al$_x$N coatings has
attracted much attention, the fundamental aspect of the elastic properties
has this far been neglected. 

This is so even though it is known that the
for example spinodal decomposition is influenced by elastic anisotropy \cite{Seol2003}
as well as the hardness enhancement observed relies on a shear modulus
difference between the formed domains \cite{Koehler1970}.
In this work we model the influence
of composition $x$ in the Ti$_{1-x}$Al$_x$N system on the elastic constants
and elastic anisotropy theoretically with first-principles density-functional
theory (DFT) and compare the obtained results with experimentally recorded anisotropy.

The good correlation between our theoretical and experimental results suggests that the
elastic properties are strongly affected by the composition, which is to date and overlooked
parameter for tailoring the properties of hard coatings.

The special quasirandom structure (SQS) method \cite{Zunger_SQS} has been applied to
model the alloys. The approximate SQS supercells were generated by optimizing the
Warren-Cowley pair short-range order (SRO) parameters \cite{Ruban2008} up to the
7th neighboring shell by applying the Metropolis-type simulated annealing
algorithm with a cost-function built from the properly weighted pair
SRO parameters. Substitutional alloys have been considered, where the
Al atoms can occupy sites only on the Ti-sublattice. Accordingly, the SRO
parameters were calculated only on the Ti-sublattice. Among the generated
SQS supercells, the ones with size of (4$\times$4$\times$3) have proved to be good
approximations of random alloys over the entire composition range. This
size means altogether 96 atoms in the supercell. The elastic constants of
the SQS supercells were obtained by first-principles total energy
calculations within DFT using the projector
augmented wave (PAW) approach \cite{Blochl1994} as implemented in the Vienna {\it ab-initio}
simulation package (VASP) \cite{Kresse1996}. The Perdew-Burke-Ernzerhof generalized
gradient approximation (PBE-GGA) \cite{Perdew1996} of the exchange-correlation
energy functional with a kinetic energy cutoff of 450 eV has been applied. Integrations
over the Brillouin zone employed a 6$\times$6$\times$6 grid of special k points.
The elastic stiffness constants were derived by finite difference technique
from the second order Taylor-expansion coefficients of the total energy.
\begin{figure}[h!]
\includegraphics[width=6.5cm]{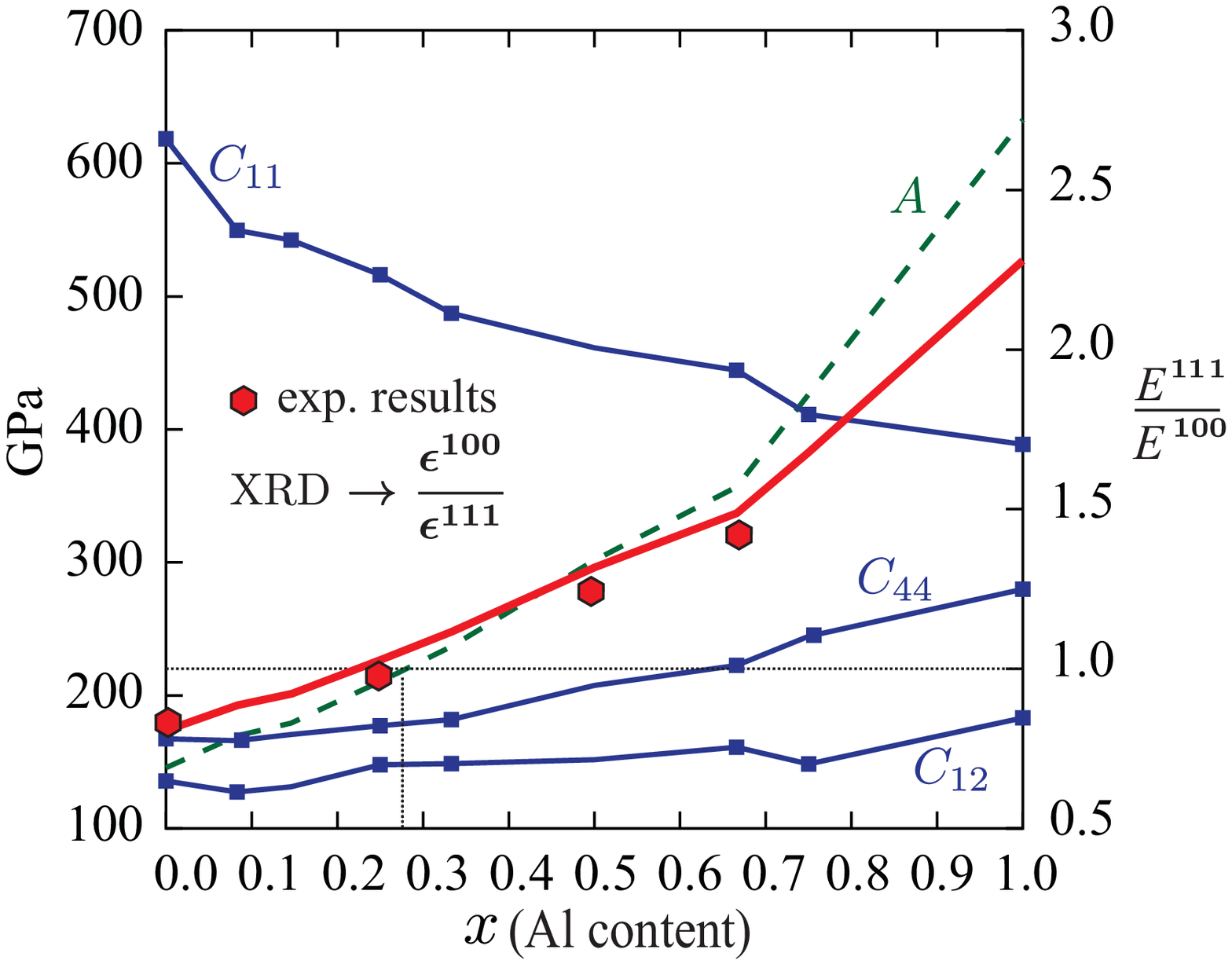}
\caption{\label{fig_elastic_constants}(Color online)
The calculated elastic stiffness constants, $C_{11}, C_{12}, C_{44}$ of Ti$_{1-x}$Al$_x$N alloys and
the extracted theoretical and experimental values of the Young's modulus ratio, $E^{111}/E^{100}$. The
dashed line shows the Zener's elastic anisotropy factor, $A=2C_{44}/(C_{11}-C_{12})$.
}
\end{figure}
Fig. \ref{fig_elastic_constants} presents
the obtained ab-initio cubic elastic constants, $C_{11}$,  $C_{12}$, and $C_{44}$ of Ti$_{1-x}$Al$_x$N
alloys as a function of Al-content. The constants change smoothly and
$C_{12}$ and $C_{44}$ increase with the amount of Al, while $C_{11}$, in comparison,
shows a pronounced decrease. The increase of $C_{44}$ has been theoretically observed for
specially chosen supercells of Ti$_{1-x}$Al$_x$N \cite{Mayrhofer2006}.

\begin{figure}[h!]
\includegraphics[width=6.5cm]{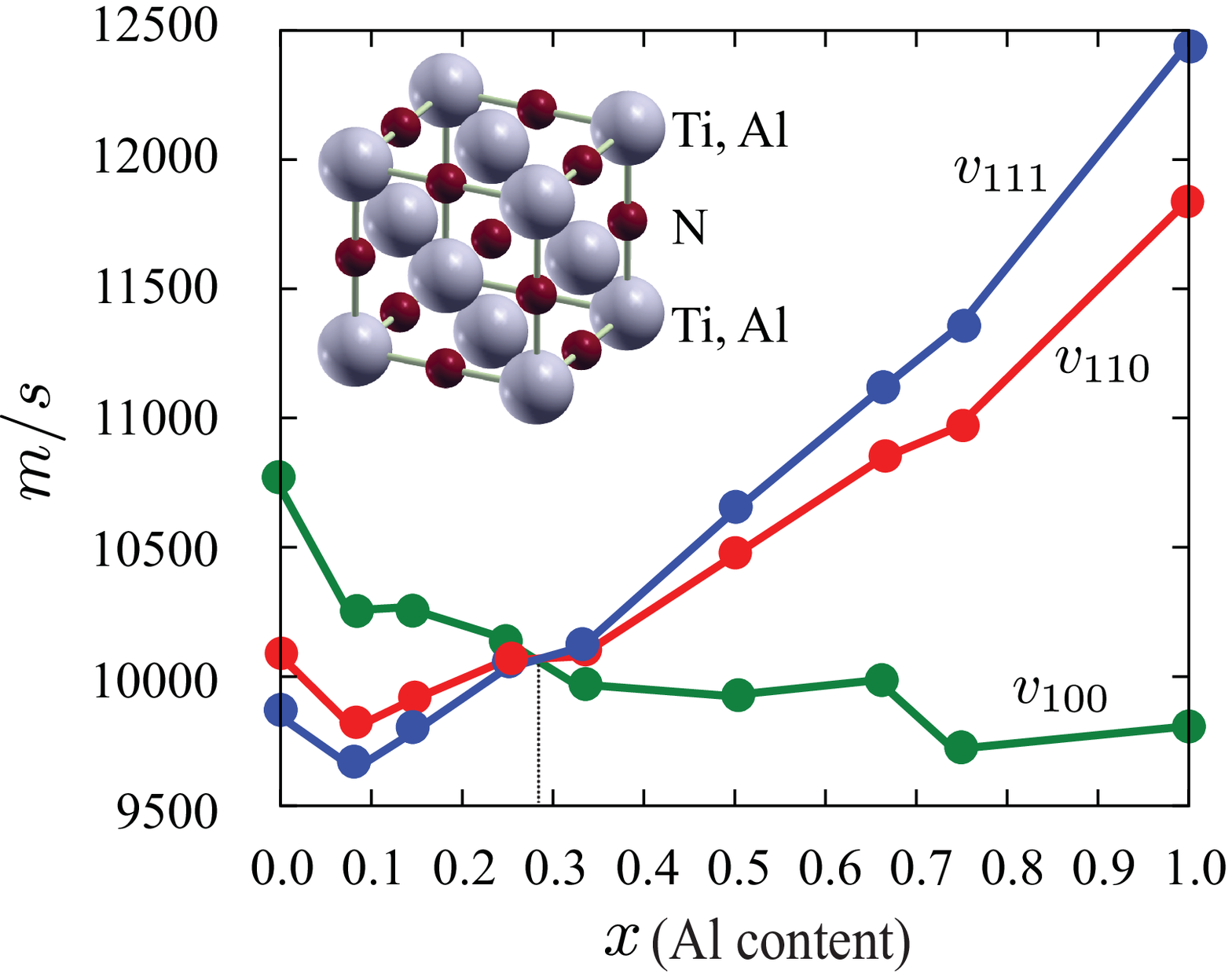}
\caption{\label{fig_sound_velocities}(Color online) Longitudinal sound velocities in directions [100], [110] and [111],
extracted from  $\rho v_{100}^2=C_{11}, \rho v_{110}^2=(1/2)(C_{11}+C_{12}+2C_{44}), \rho v_{111}^2=(1/3)(C_{11}+2C_{12}+4C_{44})$
, where $\rho$ is the density.}
\end{figure}

These facts have important implications on the general
elastic behavior of the alloy, as the longitudinal sound velocity is then
expected to decrease in the [100] direction and increase along [111]. The extracted
theoretical longitudinal sound velocities $v_{hkl}$, shown in Fig. \ref{fig_sound_velocities},
indicate elastically isotropic Ti$_{1-x}$Al$_x$N at x$\approx$0.28.
The directional anisotropy of the longitudinal sound velocities indicate
significant difference in the nature of the nearest neighbor bonds with increasing
Al content. The strong increase
of $v_{111}$ is unambiguously connected to the increasing $C_{44}$ in Fig. \ref{fig_elastic_constants}, while the
decrease of $v_{100}$ is a clear consequence of the softening of $C_{11}$, particularly at low Al content.

To quantify the elastic anisotropy in this system, we invoke Zener's elastic-shear anisotropy index,
$A=2C_{44}/(C_{11}-C_{12})$
\cite{Nye1985}.
Most of the materials show elastic anisotropy by having A$\ne$1, but in
particular cases the condition of isotropy A=1 can be fulfilled. Though
one never or only rarely expects elastic isotropy for compounds, the compositional
freedom of alloys allows for isotropic elastic behavior at a certain
concentration, here around $x\approx 0.28$, see Fig. \ref{fig_elastic_constants}.

To verify the predicted elastic behavior, arc evaporated
Ti$_{1-x}$Al$_x$N coatings with the composition
$x=0, 0.25\pm 0.02, 0.50\pm 0.02$, and $0.67\pm 0.02$ (determined by energy dispersive x-ray analysis)
were deposited. The synthesis was conducted in a 
deposition chamber [Sulzer Metaplas MZR323] using Ti$_{1-x}$Al$_x$ alloy targets, a 2Pa N$_2$
reactive atmosphere and a bias potential of -20V. The depositions were
performed at $\approx$200\,\textcelsius\ on WC/Co substrates [Seco Tools HX, chemical composition
(wt. \%) WC 94Co 6] of size $13\times 13\times 4$ mm$^3$ and hardness 1635 HV10.
The 8 $\mu$m thick coatings are dense and exhibit columnar growth that results in an isotropic microstructure
within the film plane. More details on the coating microstructure can be found elsewhere
\cite{Horling2002,Knutsson2008,Oden2009}.

Analysis was performed using 
high-energy (E=80.72 keV) synchrotron x-ray diffractometry at beamline
1-ID at the Advanced Photon Source (APS), Illinois USA. The
beam was vertically focused using refractive lenses to ~1.5 $\mu$m (full-width-half-maximum)
while the horizontal size was defined to 100 $\mu$m using slits. An ion chamber
in front of the specimen was used to measure incident beam intensity, and
a 2048$\times$2048 area detector (GE Angio) with 200$\times$200 $\mu$m$^2$ pixels was placed
2250 mm downstream from the sample. The samples were sectioned to a 1 mm
thick slice, such that diffraction data was recorded in transmission mode.
Each detector exposure consisted of Debye rings 111, 200 from the cubic B1 structure, which
were corrected for detector dark-field
current. The strains in the as deposited samples were determined through a
procedure outlined elsewhere \cite{Almer2003}. An alumina powder was used to calibrate the
detector distance and tilt angles resulting in an accuracy of the strain
determination to better than 10$^{-4}$.  The biaxial stress, $\sigma$, known to be
generated during growth \cite{Almer2001} results in strain partitioning among the grains \cite{Hedstrom2010}
depending on their crystallographic orientation. Hence,
$\sigma=E^{hkl}\epsilon^{hkl}$,
where $E^{hkl}$ and $\epsilon^{hkl}$ are the elastic (Young's) modulus and
strain of the grains contributing to the diffraction signal having a
crystallographic direction $\langle hkl\rangle$ oriented perpendicular to the growth
direction, i.e. we assume Reuss strain distribution.
The Reuss model has previously been applied to sufficiently well describe the strain distribution in
arc evaporated transition nitrides \cite{Almer2003}.
Accordingly, a measure of the elastic anisotropy can be experimentally determined
by the strain ratio
$\epsilon_{111}/\epsilon_{200}=E_{200}/E_{111}$
and since $E_{200}=E_{100}$, comparison between calculated elastic modulus
and measured strain can be made. Such comparison is given in Fig. \ref{fig_elastic_constants}
where the calculated elastic modulus ratio and measured
strain ratio are plotted as a function of Al-content. The close
agreement between measured and calculated elastic anisotropy confirms
the validity of the calculations.

\begin{figure}[h!]
\includegraphics[width=6.5cm]{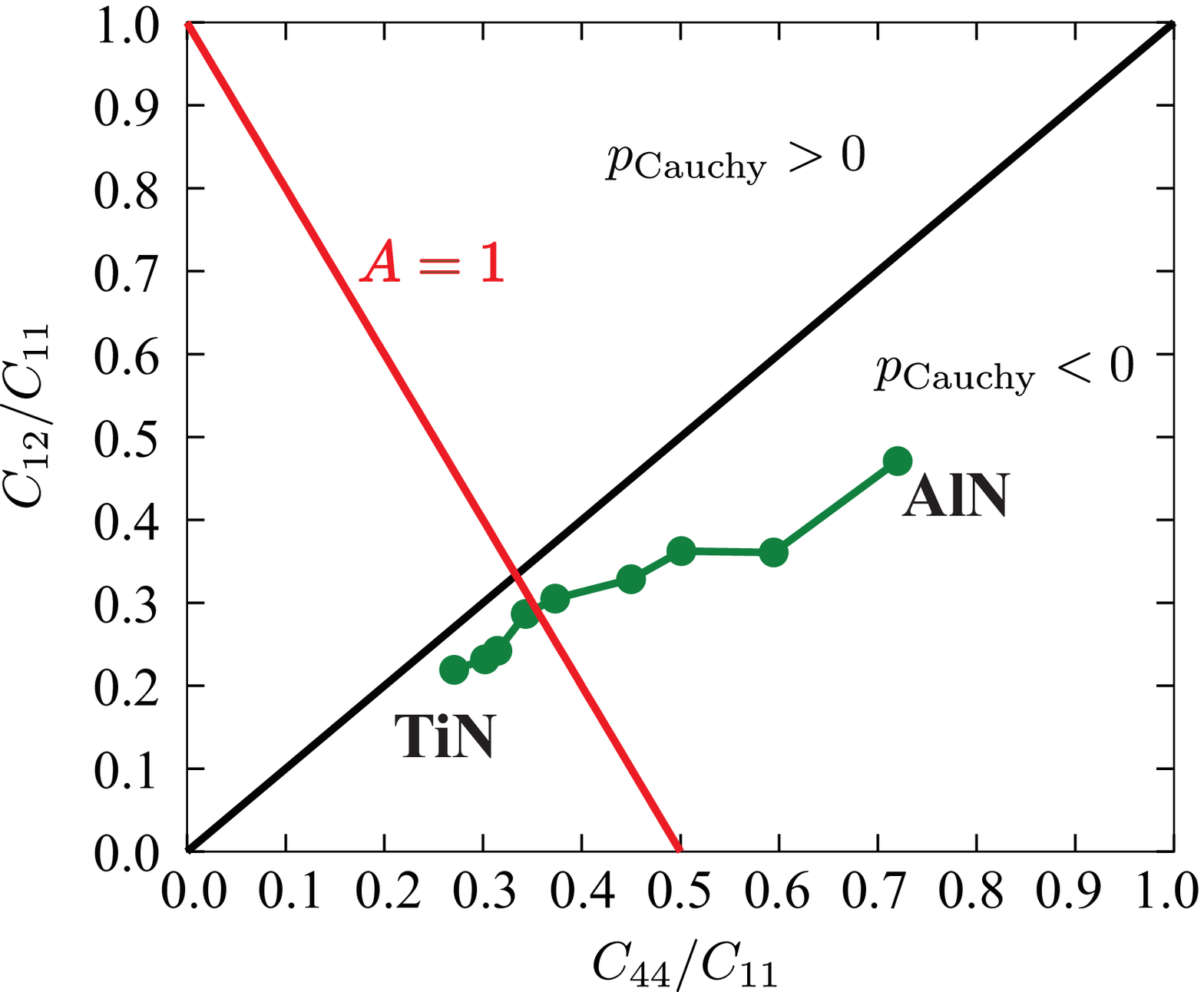}
\caption{\label{fig_blackmans_diagram}(Color online) Blackman's diagram of TiAlN alloys.}
\end{figure}

Based on {\it ab-intio} calculations of a few specially chosen supercells of TiAlN, Mayrhofer
\cite{Mayrhofer2006} extracted $C_{44}$ and suggested that its variation with Al-content was
related to the bonding in their model. To further study the interatomic
bonding in our modeling of quasi-random solid solution cubic Ti$_{1-x}$Al$_x$N alloys,
we plot the data in the form of a Blackman's diagram \cite{Blackman1938} in Fig. \ref{fig_blackmans_diagram}.
Such diagram plots the ratios of elastic constants, $C_{12}/C_{11}$ vs. $C_{44}/C_{11}$, and
reveals general characteristics of the system, such as elastic anisotropy and interatomic bonding type.
With the help of this diagram
one can determine the change in the character of bonding with composition.
The 45$^{\circ}$ solid line represents the Cauchy relationships, $C_{12}=C_{44}$ or zero
Cauchy pressure. According to Pettifor \cite{Pettifor1992} the Cauchy
pressure, $C_{12}-C_{44}$, is positive in case of fairly metallic bonds and negative
for more directional or covalent like bonds. Ti$_{1-x}$Al$_x$N alloys are all in the
covalent like region. The decreasing Cauchy pressure with increasing Al content
indicates a tendency towards stronger covalent character of the bonds, which in general results in
increased resistance against shearing and thus in increasing $C_{44}$.
The increasing longitudinal sound velocity $v_{111}$ indicates also
an increasing  covalent nature of the system. A longitudinal wave along [111] affects the 6 nearest neighbor bond
angles between the N-(Ti,Al)-N layers, as depicted in the inset of Fig. \ref{fig_sound_velocities}.
Thus, the increased $v_{111}$ results from an increased resistance to bending of these
6 bonds, which is a clear indication of the increased covalent nature of the nearest neigbour
bonds. The decrease of $C_{11}$ correlates to the decreasing bulk modulus with increasing
Al content. Thus, the more compliant nature of $C_{11}$ can be understood as weakening of
the average bond strength in the local octahedral arrangements of the B1
structure. This agrees with the obtained rapid initial decrease of $v_{100}$.
Consequently, an increased Al content changes the bonding nature of Ti$_{1-x}$Al$_x$N
alloys in two different senses. Although, the bonds become more directional,
the average strength between the octahedrally arranged atoms gets slightly weaker.
Furthermore, the observed directional anisotropy should establish an anisotropic strain
distribution in the different crystallographic directions.

The strong dependency of the elastic properties on the Al-content presented
here offers new insight to the detailed understanding of the observed age
hardening in Ti$_{1-x}$Al$_x$N alloys as an effect of spinodal decomposition.
Firstly, the nature of the Ti$_{1-x}$Al$_x$N decomposition and resulting microstructure are affected by the
elastic anisotropy, which was already pointed out by , e.g., Cahn \cite{CAHN1961}. Secondly,
as the Ti and Al-rich domains evolve during spinodal decomposition the
spatially fluctuating elastic properties will give rise to a Koehler \cite{Koehler1970}
type hardening in addition to strengthening due to the coherency strains
between the domains. Its origin is the differences in elastic shear
modulus, giving rise to repelling shear stresses acting on dislocations
gliding across the compositional gradients. The validity of this model has been
demonstrated on multilayers by for example by Chu et al. \cite{Chu1995} for several
different material systems as long as the composition modulation period
is less than approximately 20 nm. Hence, we conclude that the observed
age hardening seen for TiAlN \cite{Mayrhofer2003} and TiN/TiAlN multilayers \cite{Knutsson2010} are affected
by evolving elastic properties, and this detailed understanding can be utilized to tailor
material properties, such as hardness and thermo-mechanical response of
Ti$_{1-x}$Al$_x$N alloys.
%

This work was supported by the SSF project Designed Multicomponent coatings,
MultiFilms and the Swedish Research Council (VR). Use of the Advanced Photon
Source was supported by the U.S. Department of Energy, Office of Science,
Office of Basic Energy Sciences under Contract No. DE-AC02-06CH11357. 
Calculations have been performed at Swedish National Infrastructure for Computing
(SNIC).
%
\bibliographystyle{aps}

%
\end{document}